\documentclass[12pt,oneside,english]{article}

 \usepackage{epsfig}
\usepackage[T1]{fontenc}
\usepackage[latin1]{inputenc}
\usepackage{babel}
\usepackage{setspace}

\makeatletter
\begin{document}
\section*{ \bf Diffusion-limited reaction for the  one-dimensional trap system}
\makeatother
\bigskip \bigskip

 \centerline{\bf   D. Bar$^{a}$}
 {\bf $^a$Department of Physics, Bar Ilan University, Ramat Gan,
Israel.}
 
\bigskip
\begin{abstract}  \noindent {\it  We have previously discussed the
one-dimensional multitrap system of finite range and found the somewhat
unexpected result that the larger is the number of imperfect traps the higher is
the transmission through them. We discuss in this work the effect of a small
number of such traps arrayed along either a constant or a variable finite spatial
section. It is shown that under specific conditions, to be described in the
following,  the remarked high 
transmission may be obtained for this case also. Thus, compared to the 
theoretical large number of
traps case these results may be  experimentally  applied to real phenomena}
\end{abstract}  
\hspace{0.6 cm}  {Pacs number(s):  02.50.Cw, 02.10.Ud, 02.60.Lj}
\bigskip
\bigskip \noindent  \protect \section{Introduction \label{intr}}
\smallskip \noindent 
The   problem of diffusion through traps
\cite{Abramson,Havlin,Smol,Collins,Noyes} is generally discussed in the literature by
refering to the single trap system. In \cite{Budde} an infinite mutitrap system arranged 
over all space was discussed and the application of traps to long chain polymers was 
considered in \cite{Burlatsky}. In \cite{Bar1} the aspect of
the density of one-dimensional traps in a finite spatial section was discussed
and it was shown that the larger this density is the higher is the transmission
of particles through this dense system of traps. In this work we refer,
especially, to the small number of traps system and find the conditions under
which the transmission through them is maximal. That is, assuming an ensemble of
classical particles  we look for the conditions
that enable all or most of them   to pass through the trap system.    We 
 apply  in this work  the transfer matrix method \cite{Merzbacher,Yu} 
 used in \cite{Bar1} with
respect to the multitrap system.  \par  
In Section II we introduce the  one-dimensional
$N$  trap system, the relevant nomination and terminology and, especially, 
the appropriate transfer matrix formalism as in \cite{Bar1}. In Section III 
we  pay special attention to a small section of the $N$ trap system  that
includes $m$ traps where $m$ is a small number. We will find how the passage of
classical particles through this subsystem is influenced by the relevant
parameters of it. These parameters are the number $m$ of traps, the total extent
$L$ of the subsystem, the ratio $c$ of the total interval among its traps to the
total width of them \cite{Bar1}, the time at which the subsystem is observed and
the degree $k$ of its imperfection. Note that the ideal trap is characterized
\cite{Havlin,Smol} by $k \to \infty$ in which case all the approaching particles
are absorbed by it. \par In Section IV we discuss the behaviour 
of the transmission amplitude
for some specific values of its parameters.  
This  transmission amplitude  is defined \cite{Bar1} as the ratio
of the value of the imperfect trap coefficient  of the transmitted
particles at   the last trap  to its value at the first one. We discuss now this
transmission amplitude by considering the ratio of the ideal trap coefficient of
the density of the passing particles at the last trap to its value at the first
one. We
show that this amplitude increases to unity  for increasing  $c$ and  for  large 
values of the
remarked variables $L$, and $k$. This amplitude, on the other hand, 
 decreases with increasing $t$. Note that although one may expect the
transmission to decrease for increasing $k$ since this signifies, as remarked, a
strengthening of the ideal character of the trap system which entails a large
absorption of the passing particles,  nevertheless the obtained results are
exactly the opposite.  We show that the deviation  from the expected results is
due to the initial and boundary-value  conditions we employ for 
the imperfect trap system  discussed here (see the set (3) in \cite{Bar1} and
set \ref{e4}) here) 
which are different from those of the ideal ones 
(see the set (2) in \cite{Bar1} and set (\ref{e3}) here). That
is, the influence of the time factor that emerges from these conditions 
is much more large than that of $k$ in such
a way that it veils the expected influence of the latter especially during the
initial time.  We accompany our calculations with figures that 
exemplify the results obtained for representative values of $L$, $k$, $m$ and
$t$. 

\pagestyle{myheadings}
\markright{TRANSFER MATRIX METHOD FOR THE IMPERFECT \ldots}   
\section{ \bf Transfer matrix method for the imperfect 
 one-dimensional  trap system}
The problem we discuss here is the diffusion limited reaction in the presence of 
$N$ traps where, compared to the case discussed in \cite{Bar1}, $N$ 
   assumes  small values. These traps are arrayed in an ordered one-dimensional 
structure along a spatial axis. We denote the total width of the traps by $a$ 
and the total interval among them by $b$. Thus, for a total number of traps $N$ 
the width of each  is $\frac{a}{N}$ and the interval between any two 
neighbouring traps is $\frac{b}{(N-1)}$ since there are $(N-1)$ intervals 
among $N$ traps. An important parameter related to this system  is the ratio 
$c$ of the total interval $b$ to the total width $a$, that is,  $c=\frac{b}{a}$. 
Thus, denoting  the total length of the system $(a+b)$ by $L$ one may express the 
parameters $a$ and $b$ as \cite{Bar1} 
\begin{equation} \label{ee1} a=\frac{L}{(1+c)}, \ \ \ \  \ \ \ b=\frac{Lc}{(1+c)} 
\end{equation} 
The relevant one-dimensional initial and boundary value diffusion problem 
in the presence of $N$ traps is 
\begin{eqnarray} && \rho_t(x,t)=D\rho_{xx}(x,t), \ \ \ \ \ \ \ \ \ 0 < x \le (a+b) \nonumber \\ 
&& \rho(x,0)=\rho_0+f(x),  \ \ \ \ \ \ \ \ \ \ \ 0 < x \le (a+b) \label{ee2} \\
&& \rho(x_i,t)=\frac{1}{k}\frac{d\rho(x,t)}{dx}|_{x=x_i}, \ \ \ \ \ \   t > 0, \ \ \ 
1 \le i \le 2N,  \nonumber \end{eqnarray} 
where $\rho(x,t)$ denotes the density of the particles diffusing through the 
traps. $\rho_t(x,t)$ is the first order partial derivative with respect to the 
time variable $t$ and  $\rho_x(x,t)$,  $\rho_{xx}(x,t)$ are the first and second 
order partial derivatives with respect to the spatial variable $x$. $D$ is the 
diffusion constant which may be of two kinds;  $D_i$ and $D_e$ which are 
 the diffusion constants  inside and outside the traps respectively. 
The second and third equations in the set (\ref{ee2}) are the initial and boundary 
value conditions respectively. The range of $i$ in the third equation of the set 
(\ref{ee2}) is due to the fact that each trap has  front and back faces.  
It has been shown \cite{Bar1} that the diffusion problem from the set (\ref{ee2}) 
may be separated into the following two problems 
\begin{eqnarray} && \rho_t(x,t)=D\rho_{xx}(x,t), \ \ \ \ \ \ \ \ \ 0 < x \le (a+b) \nonumber \\ 
&& \rho(x,0)=f(x), \ \ \ \ \ \ \ \ \ \ \ \ \ \ \ \ 0 < x \le (a+b) \label{ee3} \\
&& \rho(x_i,t)=0, \ \ \ \ \ \ \ \ \ \ \ \ \ \ \ \ \ \ \ t > 0, \ \ \ \ \ 
1 \le i \le 2N,  \nonumber \end{eqnarray} 
\begin{eqnarray} && \rho_t(x,t)=D\rho_{xx}(x,t), \ \ \ \ \ \ \ \ 0 < x \le (a+b) \nonumber \\ 
&& \rho(x,0)=\rho_0, \ \ \ \ \ \ \ \ \ \ \ \ \ \ \ \ \ \ \ 0 < x \le (a+b) \label{ee4} \\
&& \rho(x_i,t)=\frac{1}{k}\frac{d\rho(x,t)}{dx}|_{x=x_i}, \ \ \ \ \ \  t > 0, \ \ \ \ \ 
1 \le i \le 2N,  \nonumber \end{eqnarray} 
The set (\ref{ee3}) is  the one-dimensional initial and boundary value diffusion 
problem in the presence of $N$ ideal traps and the set (\ref{ee4}) is that in the 
presence of $N$ imperfect traps. The general solution of the set (\ref{ee2}) is 
\begin{equation} \label{ee5} \rho(x,t)=A\rho_1(x,t)+B\rho_2(x,t) \end{equation} 
where $\rho_1(x,t)$ is the density of the ideal trap set $(\ref{ee3})$ and 
$\rho_2(x,t)$ is that of the imperfect  trap one $(\ref{ee4})$.  The 
appropriate $f(x)$ that satisfies the first and third equations of the set 
(\ref{ee3}) is $f(x)=sin(\frac{\pi x}{x_i})$. Thus, using the separation of 
variables method \cite{Renne} one may write for $\rho_1(x,t)$ and $\rho_2(x,t)$  
that satisfy the appropriate initial and boundary value conditions 
\begin{equation} \label{ee6} \rho_1(x,t)=\sin(\frac{\pi x}{x_i})
\exp(-\frac{Dt\pi^2}{x_i^2}) \end{equation} 
\begin{equation} \label{ee7}  \rho_2(x,t)=\rho_0[erf(\frac{x}{2\sqrt{Dt}})+
\exp(k^2Dt+kx)\cdot erfc(k\sqrt{Dt}+\frac{x}{2\sqrt{Dt}})] \end{equation}
The $erf(x)$ and $erfc(x)$ are the error and complementary error functions 
respectively defined as \cite{Abramov} 
\begin{eqnarray*}  && erf(x)=\frac{2}{\sqrt{\pi}}\int_0^xe^{-u^2}du \nonumber \\ 
&& erfc(x)=1-erf(x)=\frac{2}{\sqrt{\pi}}\int_x^{\infty}e^{-u^2}du \nonumber
\end{eqnarray*}
We use in the following, as in \cite{Bar1}), the transfer matrix method 
\cite{Merzbacher,Yu} an important element of which is the assumption 
that the density $\rho(x,t)$ and its partial derivative with respest to 
$x$ changes continously along the section $L=(a+b)$.   
Thus, one may equate at each of the $2N$ faces of the $N$ traps the $\rho(x,t)$ 
and $\rho_x(x,t)$ at one side of it to the corresponding quantities just at the 
other side as done in \cite{Bar1}. In such a way one may obtain $2N$ two dimensional 
transfer matrices each of them relates the values of the coefficients $A$, $B$ from 
Eq (\ref{ee5}) at one side of a face of a trap to the corresponding values of these 
coefficients at the other side of this face. We multiply together  any two 
transfer matrices related to the two faces of the same trap so as to have one 
two-dimensional transfer matrix for each trap. Thus, denoting these matrices by 
$T$ we may write the general transfer matrix equation for the one-dimensional 
$N$ trap system that relates the coefficients $A$, $B$ at the left face of the first 
trap to those at the right face of the last trap assuming that the diffusing particles 
enter the traps through their left faces       
 \cite{Bar1}). 
\begin{eqnarray}  &&
\left(
\begin{array}{c} A_{2N+1} \\ B_{2N+1} \end{array}
\right)=T(a+b)T(\frac{(N-1)(a+b)}{N})T(\frac{(N-2)(a+b)}{N})\ldots  
\label{e1} \\ && \ldots 
 T(\frac{n(a+b)}{N})T(\frac{(n-1)(a+b)}{N}) \ldots
    T(\frac{2(a+b)}{N})T(\frac{a+b}{N})\left(
\begin{array}{c} A_1 \\ B_1  \end{array}
\right) \nonumber \end{eqnarray}
Each $T$ is, as remarked,  a two-dimensional  transfer matrix 
 that relates the coeffiecients $A$ and $B$ of the ideal and imperfect trap  
density functions   of the passing particles at one side of the relevant
trap to those at the other side.    The number of
traps is  $N$ and by $n$ we denote the general trap in this system. 
Note that each  matrix $T$ depends also 
\cite{Bar1} upon the time $t$, the constant $k$  \cite{Bar1}, and upon the two diffusion constants $D_e$ and
$D_i$.  But, as seen in \cite{Bar1} (see also Eq (\ref{e1})), 
the matrices $T$ differ from each other by only the values of $x$ and share the
same values of $k$, $t$, $D_e$ and $D_i$. \par
As seen in \cite{Bar1} the element $T_{12}$ in each $T$ is always zero and the other
elements $T_{11}$, $T_{21}$  and $T_{22}$ are given, for the value of 
$x=\frac{(a+b)}{N}$, 
 by \cite{Bar1}     
\begin{equation} \label{e2} T_{11}(\frac{(a+b)}{N})=\frac{\alpha(D_e,\frac{b}{N},t)
\alpha(D_i,\frac{(a+b)}{N},t)}{\alpha(D_i,\frac{b}{N},t)
\alpha(D_e,\frac{(a+b)}{N},t)} \end{equation} 
 \begin{eqnarray} 
 && T_{21}(\frac{(a+b)}{N})=
\rho_0(\frac{\eta(D_i,\frac{(a+b)}{N},t)}
{\eta(D_e,\frac{(a+b)}{N},t)}(\frac{\xi(D_e,\frac{b}{N},t)}
{\eta(D_i,\frac{b}{N},t)} -\frac{\alpha(D_e,\frac{b}{N},t)\xi(D_i,\frac{b}{N},t)}
{\alpha(D_i,\frac{b}{N},t)\eta(D_i,\frac{b}{N},t)})) + \label{e3}  \\ &&
+\frac{\alpha(D_e,\frac{b}{N},t)}
{\alpha(D_i,\frac{b}{N},t)}(\frac{\xi(D_i,\frac{(a+b)}{N},t)}
{\eta(D_e,\frac{(a+b)}{N},t)}- \frac{\alpha(D_i,\frac{(a+b)}{N},t)\xi(D_e,\frac{(a+b)}{N},t)}
{\alpha(D_e,\frac{(a+b)}{N},t)\eta(D_e,\frac{(a+b)}{N},t)}) \nonumber \end{eqnarray} 
\begin{equation} \label{e4} T_{22}(\frac{(a+b)}{N})=\frac{\eta(D_e,\frac{b}{N},t)
\eta(D_i,\frac{(a+b)}{N},t)}{\eta(D_i,\frac{b}{N},t)
\eta(D_e,\frac{(a+b)}{N},t)},   \end{equation} 
where $\alpha$, $\xi$, and $\eta$ in the former equations are given as  
(we write them  for 
$D_e$ )  
\begin{equation} \label{e5} \alpha(D_e,x,t)=
erf(\frac{x}{2\sqrt{D_et}})+\exp(k^2D_et+kx)\cdot erfc(k\sqrt{D_et}+
\frac{x}{2\sqrt{D_et}}) \end{equation} 
\begin{equation} \label{e6} \xi(D_e,x,t)=k\exp(k^2D_et+kx)\cdot 
erfc(k\sqrt{D_et}+\frac{x}{2\sqrt{D_et}}) \end{equation} 
\begin{equation} \label{e7} \eta(D_e,x,t) =-\frac{\pi}{x}
e^{-(\frac{\pi}{x})^2D_et} \end{equation}

\bigskip 

\pagestyle{myheadings}
\markright{THE ONE-DIMENSIONAL SMALL NUMBER OF TRAPS SYSTEM}
\section{The one-dimensional small number of traps system}
 
 It was shown in \cite{Bar1} for the multitrap system that the
transmission coefficient which was calculated as the ratio of the imperfect trap
coefficient  of the
particles after passing  
 through the system to that before this  passage tends to unity for the cases 
 of:  
 1)  when the total length of the system $L=a+b$ grows. 2) when the
 total length $L$  is constant and the ratio $\frac{b}{a}$ of the total interval
 to the total width of the system increases. For these two cases the elements
 $T_{21}$ and $T_{22}$ tend to zero (as remarked, the value of the element 
$T_{12}$ is always zero) and $T_{11}$ tends to unity which are the
 required conditions to obtain  a unity value for the transmission coefficient. 
 Moreover, it has been shown \cite{Bar1} that these specific values of the
 elements   $T_{21}$,   $T_{22}$ and  $T_{11}$ are, especially, obtained in the
 limit of $N \to \infty$.    We, now, discuss, as noted, the ideal trap
 component of the transmission coefficient and find that  this kind  of  
 behaviour may
 be discerned  in  small sections of the $N$ system 
 that contain small number of traps. We show in the following 
 that the transmission
 coefficient may indeed assume, under certain conditions, a unity value  for 
 this case also. Thus, refering to a   two trap section in the  $N$ system 
  we may write,
 using Eq (\ref{e1}), the relevant matrix expression for it   
   \begin{eqnarray} 
&& \left(
\begin{array}{c} A_{2n+1} \\ B_{2n+1} \end{array}
\right)=T(\frac{n(a+b)}{N})T(\frac{(n-1)(a+b)}{N}) \left(
\begin{array}{c} A_{2(n-2)+1} \\ B_{2(n-2)+1} \end{array}
\right)=\left[
\begin{array}{cc} T_{11}(\frac{n(a+b)}{N})&0 \\ T_{21}(\frac{n(a+b)}{N}) &
T_{22}(\frac{n(a+b)}{N})\end{array} 
\right] \cdot \nonumber \\ && \cdot \left[
\begin{array}{cc} T_{11}(\frac{(n-1)(a+b)}{N}) & 0 \\ 
T_{21}(\frac{(n-1)(a+b)}{N})
&T_{22}(\frac{(n-1)(a+b)}{N}) \end{array} 
\right] \left(
\begin{array}{c} A_ {2(n-2)+1}\\ B_ {2(n-2)+1} \end{array} \right) = 
\label{e8} \\
&&=\left[
\begin{array}{cc} T_{11}(\frac{n(a+b)}{N})T_{11}(\frac{(n-1)(a+b)}{N})&0 \\ 
T_{21}(\frac{n(a+b)}{N})T_{11}(\frac{(n-1)(a+b)}{N})+
T_{22}(\frac{n(a+b)}{N}T_{21}(\frac{(n-1)(a+b)}{N})&
T_{22}(\frac{n(a+b)}{N}T_{22}(\frac{(n-1)(a+b)}{N}) \end{array} 
\right] \cdot \nonumber \\ && \cdot \left(
\begin{array}{c} A_{2(n-2)+1} \\ B_ {2(n-2)+1} \end{array}
\right) \nonumber 
 \end{eqnarray}
 
$B_ {2(n-2)+1}$ is the ideal  trap  coefficient that refers to the trap  just
before the discussed two trap section and $B_{2n+1}$ is the one that refers  to
the second trap in this specific section.  $A_{2(n-2)+1}$ and $A_{2n+1}$  are
the corresponding imperfect  traps coefficients.  The matrix 
equation (\ref{e8})  may be decomposed to yield the following expressions for
the relevant  coefficients. \begin{equation} \label{e9}   A_{2n+1}=
T_{11}(\frac{n(a+b)}{N})T_{11}(\frac{(n-1)(a+b)}{N})A_{2(n-2)+1}
\end{equation}
\begin{eqnarray} && B_{2n+1}=(T_{21}(\frac{n(a+b)}{N})T_{11}(\frac{(n-1)(a+b)}{N})+
T_{22}(\frac{n(a+b)}{N})T_{21}(\frac{(n-1)(a+b)}{N})) \cdot \nonumber \\ && 
\cdot A_{2(n-2)+1} + 
 T_{22}(\frac{n(a+b)}{N}T_{22}(\frac{(n-1)(a+b)}{N})B_{2(n-2)+1} 
\label{e10} \end{eqnarray}
Using Eqs (\ref{e8})-(\ref{e9}) we may write Eq (\ref{e10}) as 
\begin{eqnarray} && \frac{B_{2n+1}}{B_{2(n-2)+1}}=(\frac{T_{21}(\frac{n(a+b)}{N})T_{11}(\frac{(n-1)(a+b)}{N})+
T_{22}(\frac{n(a+b)}{N})T_{21}(\frac{(n-1)(a+b)}{N})}
{T_{11}(\frac{n(a+b)}{N})T_{11}(\frac{(n-1)(a+b)}{N})}) \cdot \nonumber \\ && 
\cdot   \frac{A_{2n+1}}{B_{2(n-2)+1}} + 
T_{22}(\frac{n(a+b)}{N}T_{22}(\frac{(n-1)(a+b)}{N} \label{e11} \end{eqnarray}
In order to be able to solve the last equation for 
$\frac{B_{2n+1}}{B_{2(n-2)+1}}$ we use,  for the ratio of the imperfect trap
coefficient of the last trap to the ideal one of the first, an assumption
analogous to that in \cite{Bar1}. That is, when discussing the $4NX4N$ marix
method after Eq (29) in \cite{Bar1} we assume that the larger is the number of
imperfect traps the smaller is the ideal transmission coefficient at the last
trap compared to the imperfect one at the first trap. We made here a similar
assumption for the ratio of the imperfect trap coefficient at the last trap to
the ideal one at the first. That is, the imperfect component of the density that
remains after passing all the traps must be small compared to the ideal
component of the density before approaching them.  
 We  discuss  here the case
of a small number $m$  so we may assume that this ratio depends on  
$m$ and $\frac{B_{2n+1}}{B_{2(n-m)+1}}$ as \begin{equation} 
\label{e12} \frac{A_{2n+1}}{B_{2(n-m)+1}}=
\frac{1}{(1+m^2)}\cdot \frac{B_{2n+1}}{B_{2(n-m)+1}} \end{equation} 
The last expression ensures that the ratio at the
left hand side vanishes 
  in the limit of a very large $m$ where $\frac{B_{2n+1}}{B_{2(n-m)+1}}$ tend to
  unity \cite{Bar1}.   
  Denoting the expression that multiply $\frac{A_{2n+1}}{B_{2(n-2)+1}}$ in the
first term on the right hand side of  Eq (\ref{e11}) as $c_1^{(2)}$ and the second term  
as $c_2^{(2)}$ we  write Eq (\ref{e11}), using Eq (\ref{e12}) in which $m=2$,
 as \begin{equation} \label{e13} 
\frac{B_{2n+1}}{B_{2(n-2)+1}}=\frac{c_2^{(2)}}{(1-
\frac{c_1^{(2)}}{5})} 
\end{equation}
We may generalize the last equation  that was  written for the two trap 
section for any finite number $m$ of traps so that the corresponding analog of Eq (\ref{e13}) 
is   \begin{equation} \label{e14} 
\frac{B_{2n+1}}{B_{2(n-m)+1}}=\frac{c_2^{(m)}}{(1-
\frac{c_1^{(m)}}{(1+m^2)})},  
\end{equation}
where  it may be shown that $c_1^{(m)}$ and  $c_2^{(m)}$ 
are given by the following recursive
equations \begin{equation} \label{e15}  c_1^{(m)}=  
\frac{T_{21}(\frac{n(a+b)}{N})
+T_{22}(\frac{n(a+b)}{N})c_1^{(m-1)}}{T_{11}
(\frac{n(a+b)}{N})} \end{equation} 
\begin{equation}  \label{e16}  c_2^{(m)}=c_2^{(m-1)}T_{22}(\frac{n(a+b)}{N}) 
\end{equation}
    Note that the parameter $m$ refers to the finite $m$ trap system which is
    a susbsystem of the $N$ multitrap one  whereas the parameter $n$ 
    denotes the
    general term of the last system (it actually refers to the position of the 
    last trap of the subsystem
in the larger $N$ trap system). Now, it may be shown from the definitions
    of  the variables  
    $c_1^{(m)}$ and  $c_2^{(m)}$  that  the range of $c_2^{(m)}$ that involves the
    quantities  $T_{22}$ is in the interval $(0,
    1)$ and that of  $c_1^{(m)}$ that involves $T_{11}$ and    
    $T_{21}$ is in $(-\infty, +\infty)$. Also, it may be seen   
    that  $c_1^{(m)}$ grows in absolute value with $m$ and  $c_2^{(m)}$
    decreases to zero with increasing $m$  so that in the limit of very large
    $m$ the transmission amplitude tends to unity as shown in \cite{Bar1}. 
    The same result is obtained also for increasing  $c$ where  $c_1^{(m)}$ 
    decreases to zero and  $c_2^{(m)}$ increases to its maximum value of unity
  so that in the limit of very large $c$ the transmission amplitude tends to
  unity as may be seen from Eq (\ref{e14}) (see also  \cite{Bar1}).
  
  \pagestyle{myheadings}
\markright{RESULTS FOR SOME GIVEN VALUES OF THE TRAP SYSTEM \ldots}     
     \section{ \bf  Results for some given values of  
    the  trap system parameters $m$, $L$, 
    $k$, and $t$  }
    Unlike the discussion in \cite{Bar1} which, especially, takes account of
    a large number of traps  we discuss here, as remarked,  the influence of a small number
    of them. 
      First of all  one 
    finds, as expected, that the smaller is the number of traps $m$ the easier
    is for the classical particles to pass through them. The criterion for this
    transmission  is, as remarked    the ratio 
     of the value of the ideal trap coefficient at the last trap (of the $m$
     member subsystem)  to its
     value at the first one. This is the ratio $\frac{B_{2n+1}}{B_{2(n-m)+1}}$ 
      from Eq (\ref{e14}) 
       which may 
    be regarded as a transmission amplitude. Note that this amplitude may have
    values  outside the range of $(0, 1)$.  An easy passage through the trap 
    system is obtained not only for small values of $m$ but also for large 
    values of the total length $L$ of the system as we have found in \cite{Bar1} 
    for the multitrap case. The same result is obtained also for 
    large $k$.  
      Note that the nature of the change of the transmission amplitude 
      with time is opposite to that regarding $k$ and $L$. That is, this
      amplitude decreases with increasing time. Moreover, this decrease occurs
      in a very fast manner, especially at the initial time, 
       compared to the remarked increase with $k$ and $L$ as may be seen in
       Figures 6-8.     
       \par
     Each of the following eight figures  contains  six curves of the
     transmission amplitude from Eq (\ref{e14}) as functions of the parameter
     $c$ for the six values of $m=1, 3, 5, 7, 9, 11$. As remarked and shown in
     \cite{Bar1} with respect to the multitrap system 
     this transmission amplitude tends to unity for large values of
     $c$. We find here the dependence of the transmission upon $c$ for small $m$
     and the small range of   $0.001 \le c \le 20$.  All the curves in the eight 
     figures are drawn for the specific values of $D_e=0.5$ and $D_i=0.1$. These
     values yield results that are qualitatively similar for a wide class of
     different applications that use the trap system as a model (for example,
     $0.5 \frac{cm^2}{sec}$ is the order of magnitude one may find in the
     literature for the diffusion constant $D$ at room temperature and
     atmospheric pressure (p. 337 in \cite{Reif})).    
    We find that the larger is $m$ the slower is the approach of its 
    corresponding curve to
    unity compared to that of the smaller $m$ curves. Thus,   
    not all the six curves for $m=1, 3, 5, 7, 9, 11$ are actually shown in each
    figure as in Figures 1 and 4 in which the larger $m$ curves are merged
    with the abcissa axis.  The  correct order of the curves in each figure   
      is downward so that smaller  values of 
    $m$ fit
    the upper curves (the graph for $m=1$ is the upper one, that for $m=3$ is
    the second from above and so on).  \par
    The  group of  Figures 1-8  
       demonstrate  this behaviour of the transmission amplitude from
     Eq (\ref{e14}) as function of $c$. The first  three figures, each
     composed of six curves for $m=1, 3, 5, 7, 9, 11$,  show how the
     transmission amplitude changes with $c$ for the same values of $k=1$ and
     $t=1$ but three different values of $L=5, 13, 27$.   
      Figure 1   is drawn for $L=5$ and shows only the curves for 
     $m=1, 3, 5, 7$ whereas those for $m=9, 11$ are merged with the 
     abcissa axis. 
         Figure 2,  which  is for $L=13$, shows all
     the six curves approaching unity as $c$ grows but, as remarked, 
      this approach is slower the larger is $m$.  Figure 3,  
       which is drawn for $L=27$, shows once again all the six curves
      approaching unity  but now even the larger $m$-values 
      curves tend to unity already at small values of $c$ compared to 
        Figures 1-2. Thus, as remarked and as shown for the multitrap
       system in \cite{Bar1}, the approach
       of the transmission amplitude to unity is more apparent and faster,  even
       for small $c$,  the larger is $L$. \par 
       Figures 4-5  show how the
       transmission amplitude as a function of $c$ changes with $k$. Each curve
       from the total six curves of each figure in the group of Figures 4-5 
        is drawn for
       $L=5$ and $t=1$.  Figure 4 is for $k=5$ and one may see
       only the  curves for the 4 smaller values of $m$ that tend to unity for
       increasing $c$. The other two curves for $m=9, 11$ are merged with the
       abcissa axis.  
        Figure 5, which  is drawn for $k=27$,   shows now all the six curves 
        approaching   unity for increasing values of $c$. 
       Thus, as remarked,  the higher $k$ values
       guarantees an easy transmission of the passing particles through the
       system. Note that, as remarked, this high transmission 
        for increasing $k$ is contrary to what one may expect that  large
       $k$ entails a large absorption \cite{Havlin,Smol,Bar1} 
       of the passing particles. 
       The deviation of
       the obtained results from the expected ones is because the trap problem
       we try to solve here, as in \cite{Bar1}, is the imperfect trap one and
       not the ideal one. Thus, the initial and boundary-value 
        conditions employed are not the ideal ones (the set (3))  
	but the imperfect (the set (4)) and these 
	may cause a large transmission even at the ideal trap limit of $k \to 
	\infty$ as actually shown in \cite{Bar1} (see Figure 2 there). The 
	presence of the time factor in the initial and boundary-value conditions
	introduces interesting results that do not appear in the absence of it.
	For example, the analogous quantum one-dimensional multibarrier system 
	along a finite section \cite{Bar2} does not involve any time variation 
	and as a consequence the kind of change with time found here 
	 is not encountered there \cite{Bar2}. This kind of change is
	especially realized in the much more apparent and conspicuous manner,
	compared to that encountered for  $L$ and $k$, by which the
	transmission amplitude as function of $c$ changes  for different
	values of the time $t$. First of all, unlike the cases for $k$ and $L$,  
	 this amplitude decreases with
	increasing $t$ especially at the initial values of it. 
	 This is seen  in 
	Figures  6-8 where all the six curves in each figure  is drawn
	for $L=13$ and $k=7$. The curves of  Figure 6 are
	graphed for $t=0.01$ and one may see that all the six curves tend
	uniformly as a single graph to unity already at small values of $c$. 
	Figure 7  is drawn for $t=1.6$ and one may see how at the  small 
	time span of 1.59 the curves
	become widely separated from each other 
      so as  those that correspond to the higher $m$ values 
	 tend slowly to unity compared to those of the lower $m$. 
	 This form of the figure generally  remains stabilized with 
	time and  change only slightly by further increasing the time. In other
	words, a very large change  in the behaviour of the transmission
	amplitude, as a function of $c$, occurs during the initial time and then
	it remains almost stabilized. 
	To further demonstrate the large
	influence of time we show in Figure 8 the transmission amplitude,  
	as  
	function of $c$, for $L=13$, $k=7$ and $t=1$ and for $m=1, 3, 5, 7, 9,
	11$. A very similar  figure 
	is shown in Figure 2 which is drawn  for the same values of $L$, $t$ and
	$m$ but for $k=1$. 
	That is, increasing
	$k$ from $k=1$ by 6 units,  keeping the same values of 
	$L=13$ and   $t=1$,  
	have  a negligible influence upon the transmission amplitude. 
	But increasing the time by only 0.6,  keeping the former  values of $L$, 
	 $m$ and
	$k$,  results in a discernable effect upon the
	transmission amplitude as shown in Figure 7 which is drawn for the same
	values of $L$,
	$k$ and $m$ as in Figure 8 but at $t=1.6$ (compare the two figures 
	7-8).               
 \bigskip \noindent \protect \section{\bf  Concluding Remarks}
We have discussed in this work the effects of a one dimensional 
trap system upon the density of the passing classical particles. We have limit
our discussion to the case of small number of traps (the large number case was
discussed in \cite{Bar1}). As our
analytical means we use the transfer matrix method discussed in \cite{Bar1} 
with
respect to the one-dimensional multitrap system. We have shown that the
transmission amplitude tends to unity,  for growing $c$,  
not only in  the
limit of a very large number of traps as in \cite{Bar1} but also, under certain
conditions, for the small number of them. These conditions involve either a
large value of the parameter $k$  or  of the total length  $L$ of the system. These results have been
exemplified for specific values of $k$ and $L$ and demonstrated by the attached
figures. \par  Unlike the remarked 
change of this amplitude with respect to $k$ and $L$ it has an opposite
behaviour regarding the time $t$. That is, it decreases for all values of $m$  
as $t$ 
increases where this decrease is larger for large $m$.   Also, compared to $k$
and $L$,     this  change with time  is very fast 
 especially at the initial  time and then the transmission amplitude stabilizes 
 and changes only slightly with time.     
 We have also  shown for
 small $m$, as  for the multitrap system in \cite{Bar1}, that the 
 imperfect
 character of the system which is expressed in its initial and boundary-value
 conditions causes it to behave contrary to what is expected for large $k$. 
 That is, although large value of $k$ indicates, as remarked, a large absorption
 of the passing particles,  nevertheless,   we find a high transmission for 
 large $k$ due to
 the appearance of time in the initial and boundary-value conditions. The large
 influence of the time upon the transmission amplitude have been shown and
 demonstrated in Figures 6-8.
 
 \noindent \protect \section*{\bf Acknowledgement }
  \bigskip  \noindent  I wish to thank S. A. Gurvitz for discussions on this 
subject 

 \bigskip \bibliographystyle{plain}

\begin{figure}
\begin{minipage}{.48\linewidth}
\centering\epsfig{figure=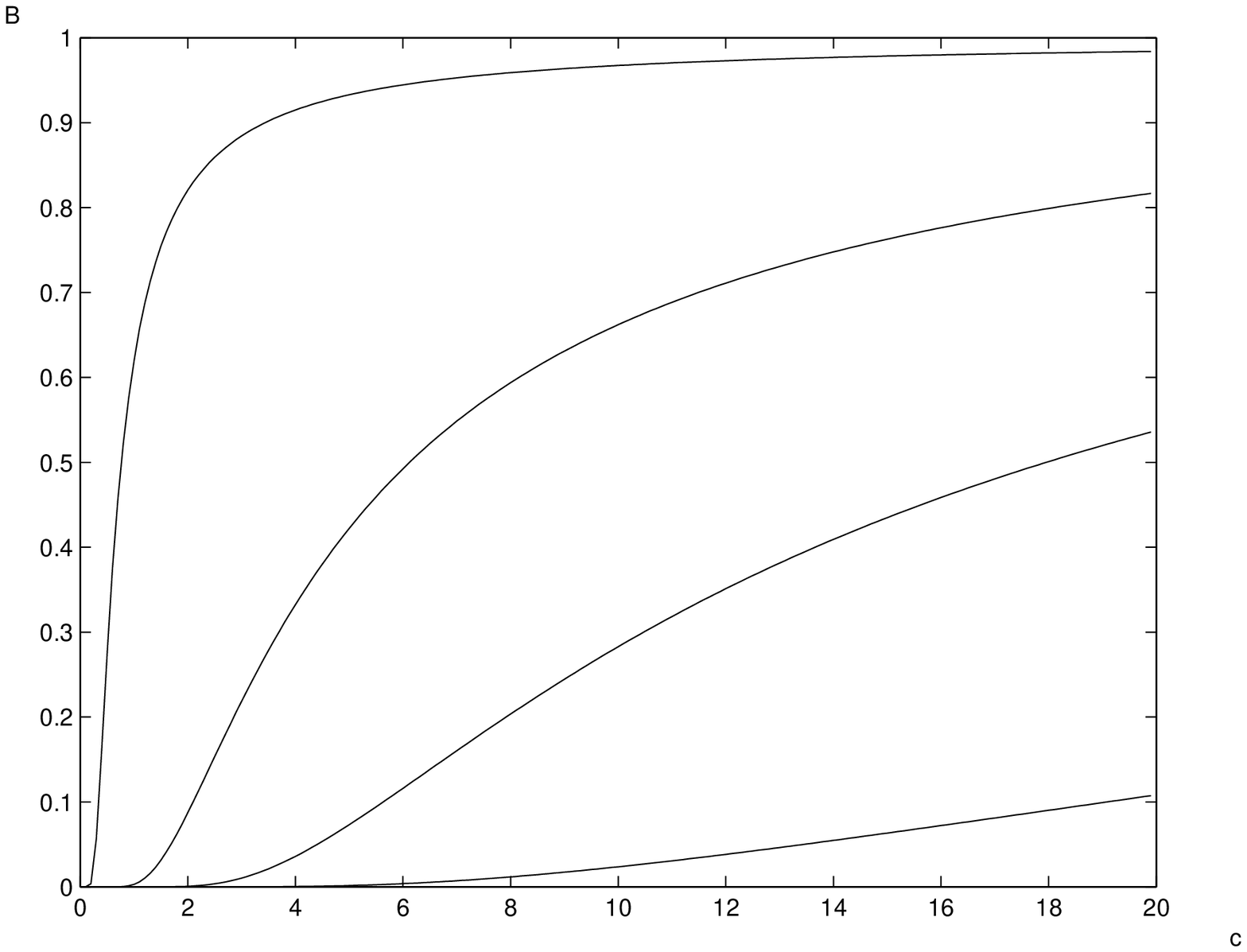,width=\linewidth}
\caption{The 4 curves show the transmission amplitude from Eq (\ref{e14}) as 
functions of the ratio $c$  
for  $m=1, 3, 5, 7, 9, 11$. Note that the curves for the larger $m$ values 
merge with the abcissa axis and are not shown. All the curves are 
drawn for the same values of $L=5$, $k=1$  $t=1$ and they all tend to unity 
for large values of $c$.   As seen,  the larger is $m$ the slower is its approach
    to unity compared to that  of the smaller $m$.}  

\end{minipage}  \hfill
\begin{minipage}{.48\linewidth}
\centering\epsfig{figure=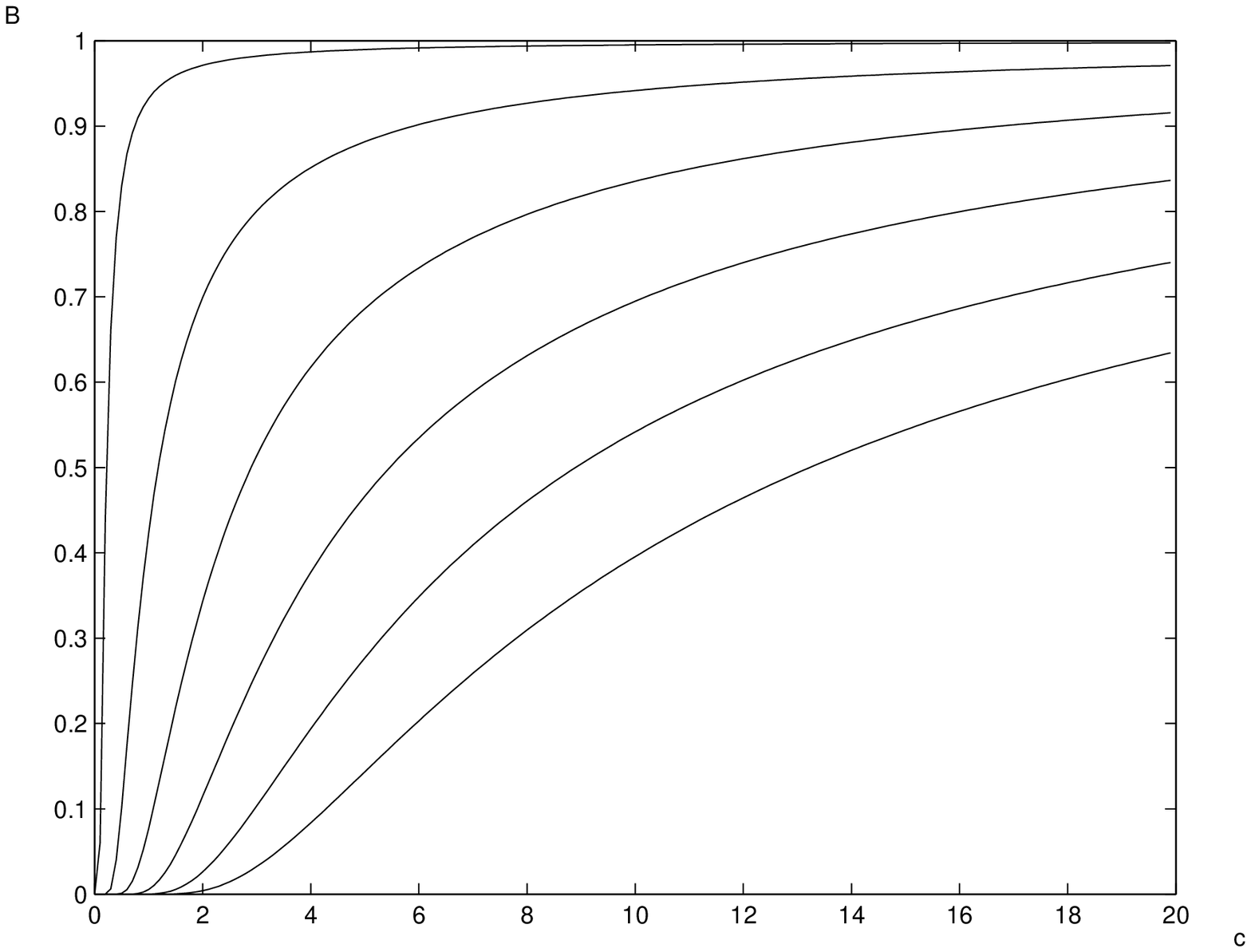,width=\linewidth}
\caption{The 6 curves show the transmission amplitude from Eq (\ref{e14}) as 
functions of the ratio $c$ for exactly the same values of $m$, $k$ and $t$ as 
those 
of Figure 1 but for $L=13$. Note that due to the larger $L$ value 
 all the 6 curves are shown (compare with Figure 1).  As in Figure 1 
 (and all the other 
figures of this work)  all the curves tend 
to unity for large $c$ where the approach to unity is slower for the larger $m$ values.}
\end{minipage}

\hspace{3.5 cm}
\begin{minipage}{.48\linewidth}
\centering\epsfig{figure=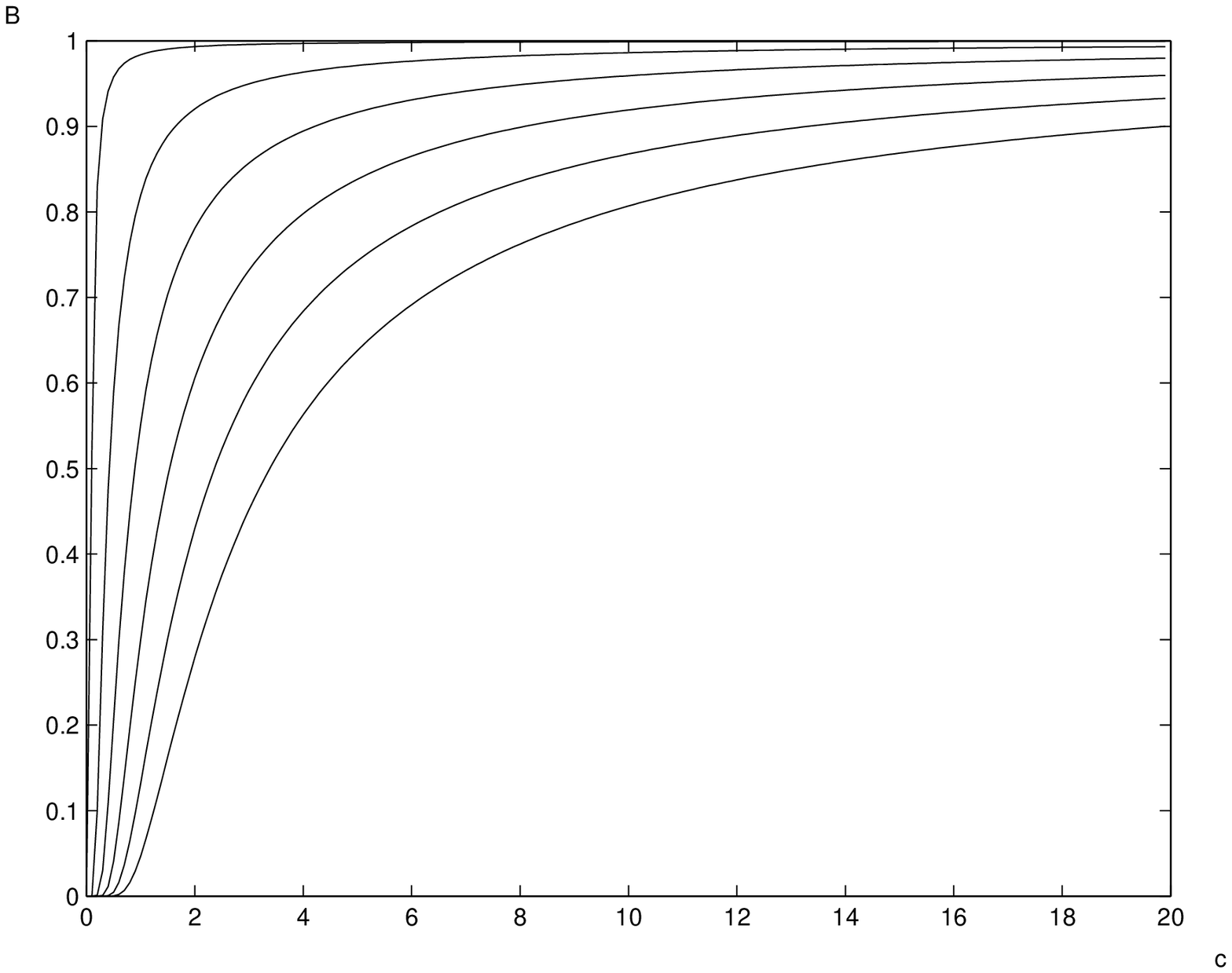,width=\linewidth}
\caption{The curves here are drawn under exactly the same conditions and for the 
same values od $m$, $k$ and $t$ as those of Figures 1-2 except that $L=27$. 
Comparing this Figure to the former two Figures one realizes that the curves 
approach unity not only for large $c$ but also for large $L$. }
\end{minipage}
 \end{figure}

\newpage 

\hbox{}

\bigskip \bigskip \bigskip \bigskip \bigskip \bigskip \bigskip 

\begin{figure}
\begin{minipage}{.48\linewidth}
\centering\epsfig{figure=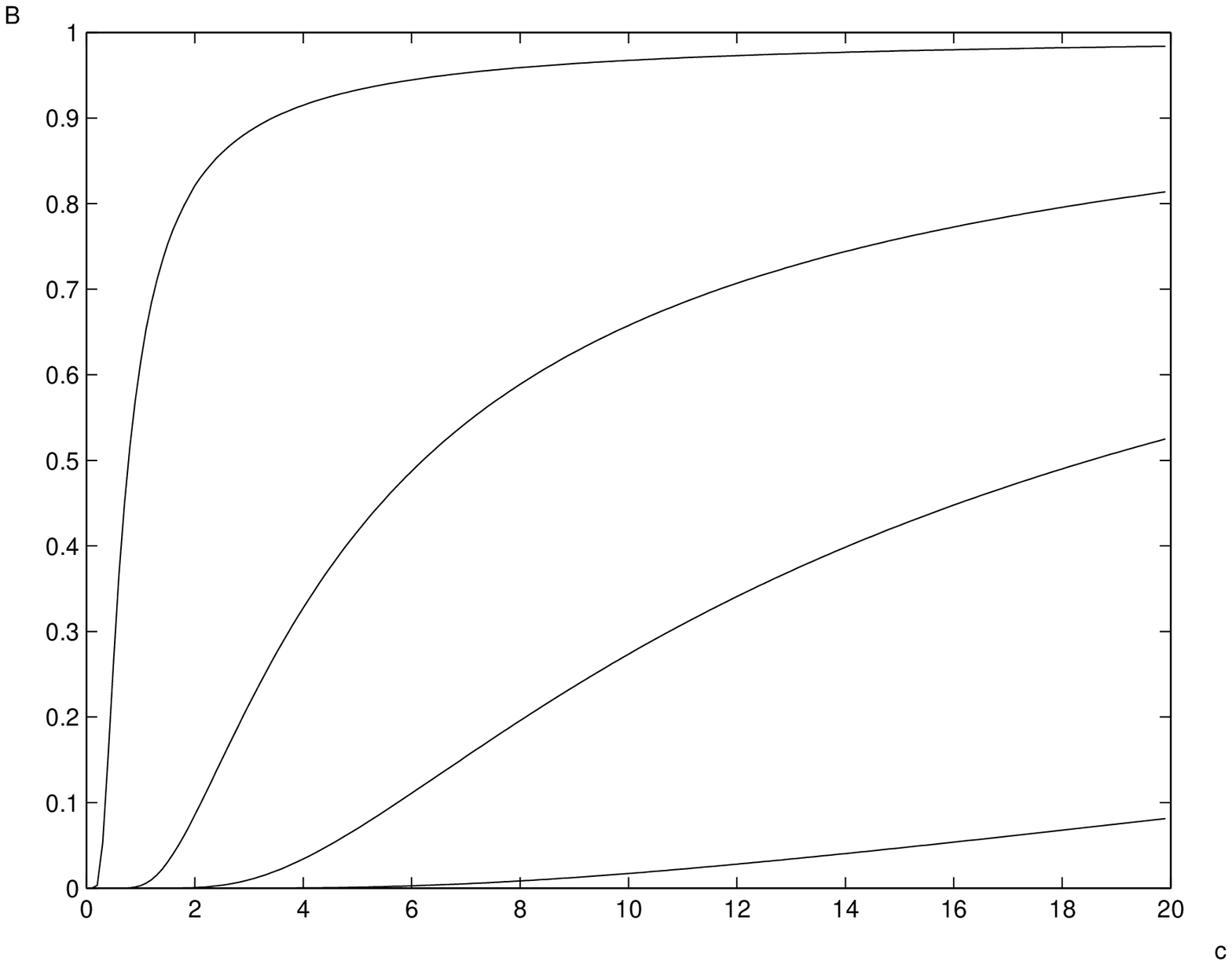,width=\linewidth}
\caption{ The 4 curves show the transmission amplitude from Eq (\ref{e14}) as 
functions of the ratio $c$  
for  $m=1, 3, 5, 7, 9, 11$. As in Figure 1  the curves for the larger $m$ values 
merge with the abcissa axis and are not shown. All the curves are 
drawn for the same values of $L=5$, $k=5$  and $t=1$ and they all tend to unity 
for large values of $c$ where  this approach is slower for larger $m$.} 
 
\end{minipage}  \hfill
\begin{minipage}{.48\linewidth}
\centering\epsfig{figure=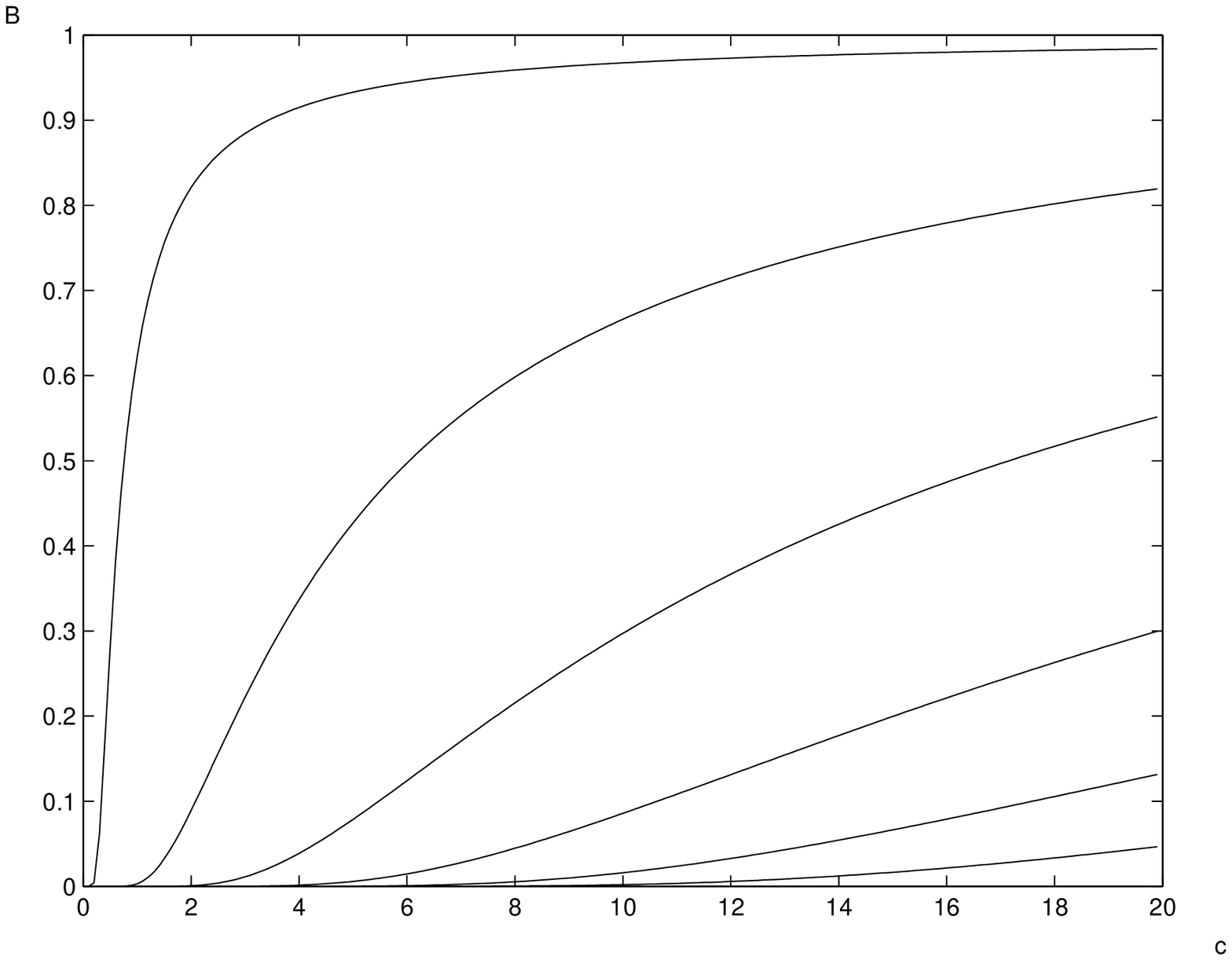,width=\linewidth}
\caption{The curves here are drawn under exactly the same conditions and for the 
same values od $m$, $L$ and $t$ as those of Figure 4 except that $k=27$. 
Note that due to the larger $k$ value 
 all the 6 curves are shown (compare with Figure 4).  
Comparing this Figure to the former four Figures one realizes that the curves 
approach unity not only for large $c$ and $L$ but also for large $k$. }
\end{minipage}
 \end{figure}
\newpage

\begin{figure}
\begin{minipage}{.48\linewidth}
\centering\epsfig{figure=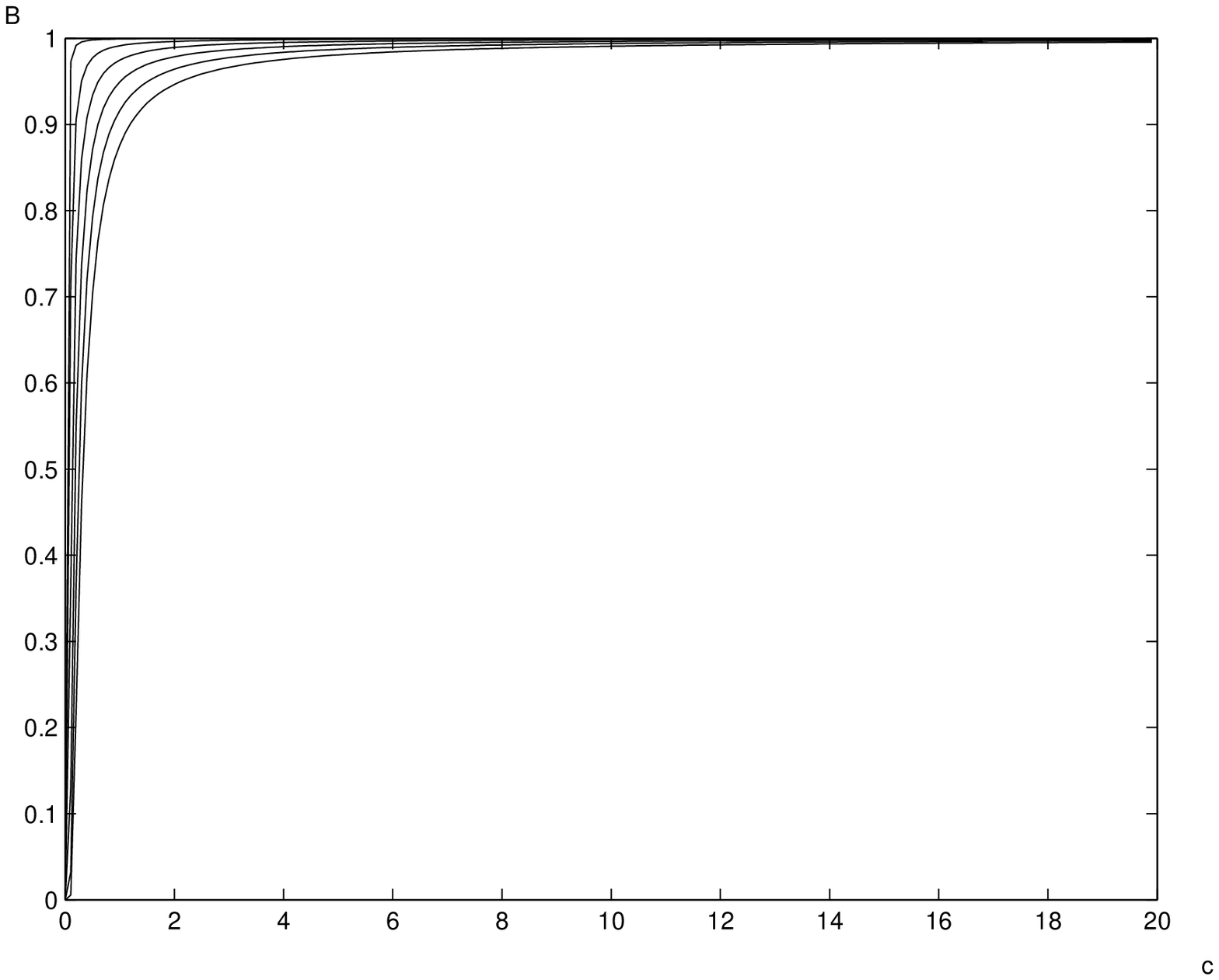,width=\linewidth}
\caption{The 6 curves show again the transmission amplitude from Eq (\ref{e14}) 
as functions of $c$  for the values of $m=1, 3, 5, 7, 9, 11$ and  for $L=13$, 
$k=7$ 
and $t=0.01$. Compared to the former 
figures one sees that for this small value of $t$ all the curves approach almost  
immediately and together to unity. That is, for small $t$ the 6 curves do not
differ much from each other.  }  

\end{minipage} \hfill
\begin{minipage}{.48\linewidth}
\centering\epsfig{figure=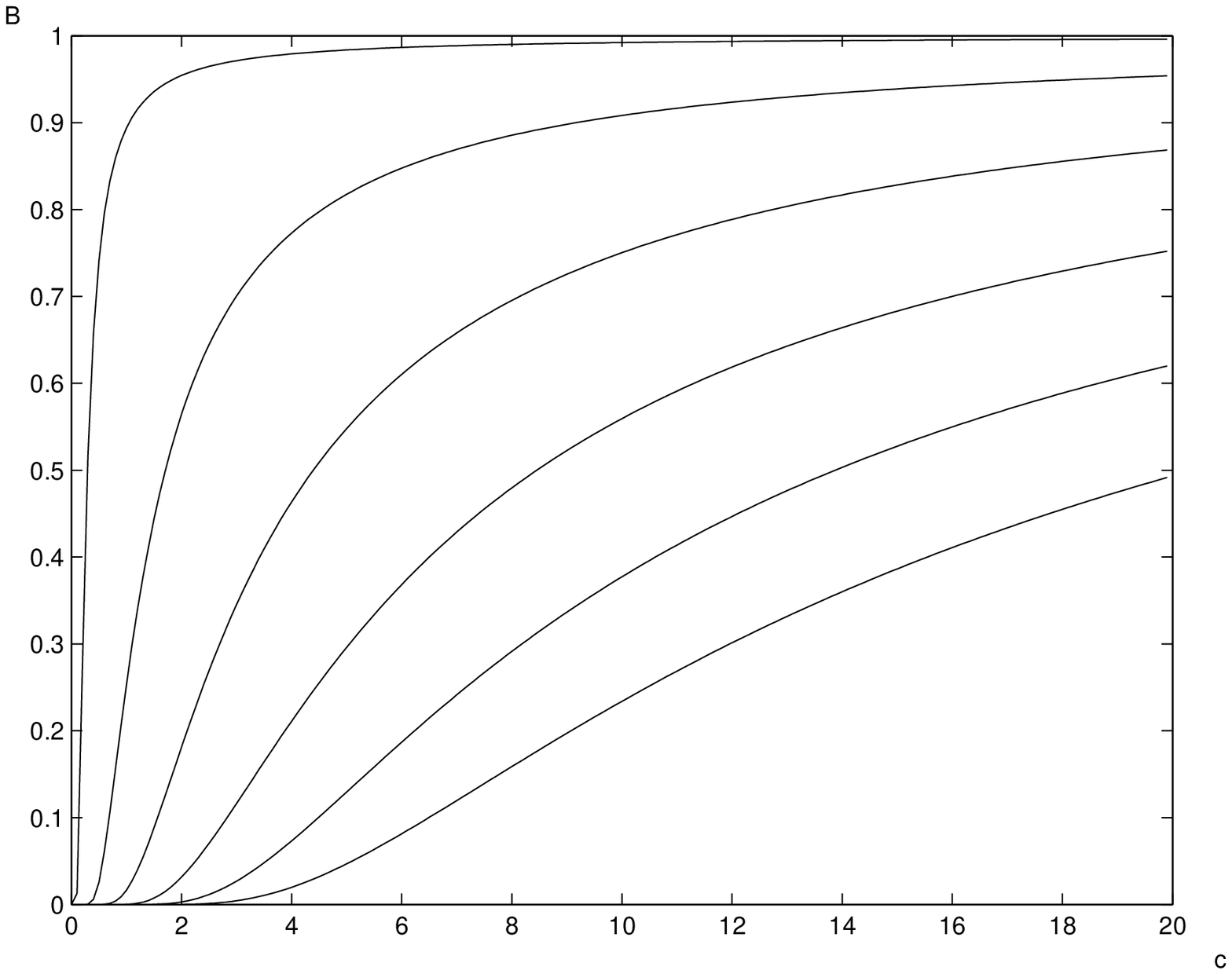,width=\linewidth}
\caption{The curves here are drawn under  the same conditions and for the 
same values of $m$, $L$ and $k$ as those of Figure 6 except that $t=1.6$. 
Note the large change 
     caused to the tramnsmission amplitude by slightly  increasing the
     time (compare with Figure 6). Also note that the
     character of the change with $t$ is opposite to that with $k$ and $L$, that
     is, the transmission amplitude  decreases with increasing time.}
\end{minipage}

\hspace{3.5 cm}
\begin{minipage}{.48\linewidth} 
\centering\epsfig{figure=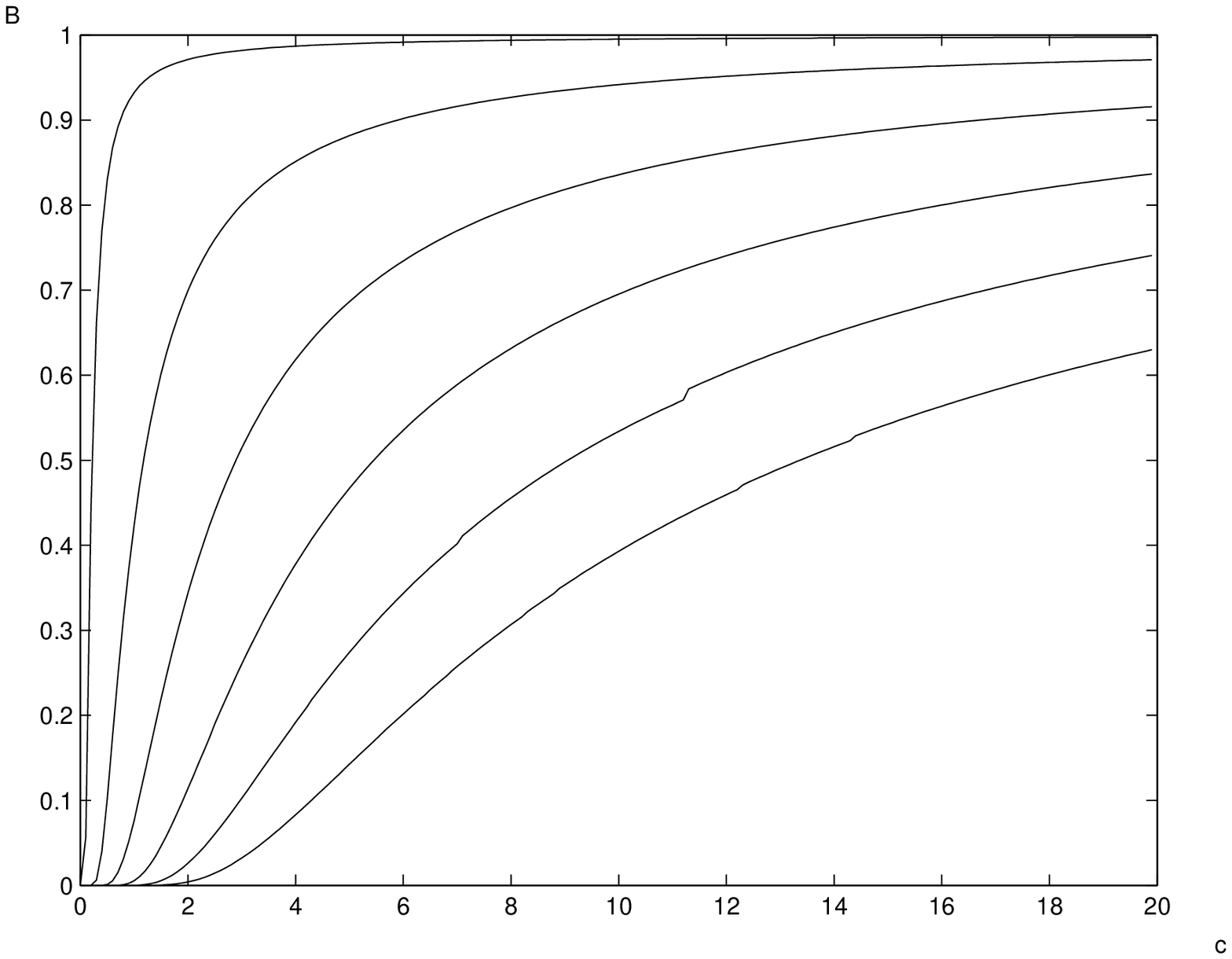,width=\linewidth}
\caption{Figure 8  is drawn for the same values of 
$L$, $k$  and $m$  as those of Figure 7 but for $t=1$. Comparing  Figure 8 with 
Figure 2 which was drawn for the same values of $L$ and $t$ but 
for  $k=1$  one realizes that increasing $k$ by 6 units, keeping the same values
of $L$ and $t$, 
has very
little influence upon the transmission amplitude. But  increasing 
the time from the value it has in this figure    by only 0.6, 
keeping the same values of $L$ and $k$   
 results in an apparent difference (compare Figures 7-8).}

\end{minipage}

\end{figure}

\end{document}